\documentclass[twocolumn,english]{article}
\usepackage[T1]{fontenc}
\usepackage[latin9]{inputenc}
\usepackage{refstyle}
\usepackage{units}
\usepackage{amsbsy}
\usepackage{amstext}
\usepackage{graphicx}

\makeatletter

\AtBeginDocument{\providecommand\figref[1]{\ref{fig:#1}}}
\RS@ifundefined{subsecref}
  {\newref{subsec}{name = \RSsectxt}}
  {}
\RS@ifundefined{thmref}
  {\def\RSthmtxt{theorem~}\newref{thm}{name = \RSthmtxt}}
  {}
\RS@ifundefined{lemref}
  {\def\RSlemtxt{lemma~}\newref{lem}{name = \RSlemtxt}}
  {}

\date{}

\makeatother

\usepackage{babel}
\begin{document}
\twocolumn[
\begin{@twocolumnfalse}

\title{Deep Neural Network Probabilistic Decoder for Stabilizer Codes}

\author{Stefan Krastanov$^{1}$, Liang Jiang$^{2}$}

\maketitle
{\small{}$^{1}$Department of Physics, Yale University, Yale Quantum
Institute}{\small \par}

{\small{}$^{2}$Department of Applied Physics, Yale University, Yale
Quantum Institute}{\small \par}

{\small{}Correspondence to liang.jiang@yale.edu}{\small \par}
\begin{abstract}
Neural networks can efficiently encode the probability distribution
of errors in an error correcting code. Moreover, these distributions
can be conditioned on the syndromes of the corresponding errors. This
paves a path forward for a decoder that employs a neural network to
calculate the conditional distribution, then sample from the distribution
- the sample will be the predicted error for the given syndrome. We
present an implementation of such an algorithm that can be applied
to any stabilizer code. Testing it on the toric code, it has higher
threshold than a number of known decoders thanks to naturally finding
the most probable error and accounting for correlations between errors. 
\end{abstract}
\end{@twocolumnfalse}
]

\begin{figure*}
\centering{}\includegraphics[width=17cm]{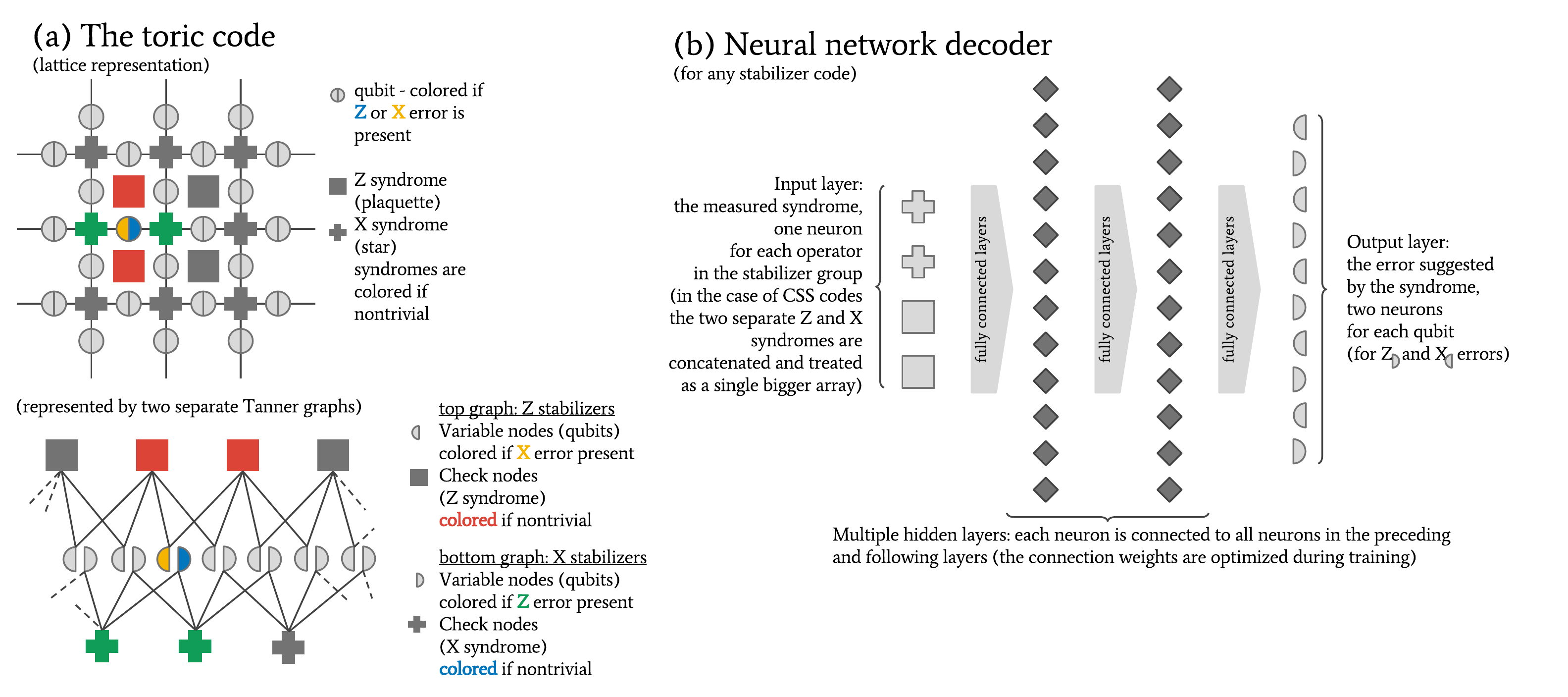}\caption{\textbf{\label{fig:ToricCodeDecoder}Quantum Error Correcting Codes.}
A very general class of QEC codes is the class stabilizer codes, defined
by the stabilizer subgroup of the physical qubits that leaves the
state of the logical qubits unperturbed. Our neural architecture can
be readily applied to such codes, however many codes of practical
interest (like the one we are testing against) have additional structure
that would be interesting to consider. The example in (a) shows a
small patch of a toric code, which is a CSS code (the stabilizer operators
are products of only Z or only X operators, permitting us to talk
of Z and X syndromes separately). Moreover, the toric code possesses
a lattice structure that provides for a variety of decoders designed
to exploit that structure. Our decoder, depicted in (b), does not
have built-in knowledge of that structure, rather it learns it through
training. Due to size constraints, the depictions present only a small
subset of all qubit or syndrome nodes.}
\end{figure*}
Constructing a physical computing machine, whether a classical or
a quantum one, requires, inescapably, the implementation of an error
correcting mechanism that guards against the noise picked up from
the environment and the imperfections in the operations being performed
\cite{Shannon48,Lidar13,Terhal15}. Early in the development of both
classical and quantum computers, ``threshold'' theorems were proven
to show the existence of encoding schemes which reliably store and
process information in a ``logical'' set of bits (or qubits), by
encoding it redundantly on top of a bigger set of less reliable ``physical''
bits (or qubits), as long as the error rate on the physical layer
is smaller than a fixed threshold \cite{von1956probabilistic,NC00}.
The vast majority of quantum error correcting codes fall in the class
of stabilizer codes (a generalization of the classical linear codes)
\cite{gottesman1997stabilizer}. They are characterized by the group
of stabilizer operators that preserve the logical states (similarly
to the list of constraints represented by the parity check matrix
H for classical linear codes). The list of nontrivial stabilizer operator
measurements (or violated parity constraints for a classical code)
is called the syndrome of the error. While providing for efficient
encoding, linear and stabilizer codes do not necessarily have known
efficient decoding algorithms that can deduce from a given syndrome
what errors have occurred.

In the general case decoding a stabilizer code is an NP-hard problem.
An active area of research is the design of codes with some additional
algebraic structure that permits efficient decoders, but still retains
high rates (ratio of logical to physical qubits) with acceptable distances
(maximal number of correctable errors on the physical qubits). Schemes
like the CSS approach \cite{calderbank1996good,Shor96,steane1997active}
permit the creation of quantum codes from classical codes, but they
do not guarantee that the decoder that worked for the classical code
still works for the quantum one. A particularly interesting example
is the class of LDPC codes \cite{gallager1962low,MacKay96} which
are high-performing classical codes with efficient decoders, however
those decoders do not work for the quantum LDPC codes \cite{poulin2008iterative}.

Here we present a decoding algorithm that can be applied to any stabilizer
code \textemdash{} the decoder employs machine learning techniques
to ``learn'' any structures that would make the approximate decoding
problem easier than the general NP-hard decoding problem: it ``learns''
the probability distributions of errors conditioned on a given syndrome
and efficiently uses samples from that distribution in order to predict
probable errors. The conditional probability distribution is encoded
in a deep neural network. The ``learning'' involves training the
neural network on pairs of errors and corresponding syndromes (generated
from an error model for the physical qubits and a parity check matrix
for the code in use). We test the algorithm on the toric code (\figref{ToricCodeDecoder}a)
definied on a two-dimensional lattice on a torus \cite{Kitaev03}.
Since the toric code has low-weight local stabilizers, it is also
a quantum LDPC code with structure that impedes typical belief propagation
algorithms. Our decoder significantly outperforms the standard ``minimal-weight
perfect matching'' (MWPM) decoder \cite{dennis2002topological,Edmonds65}.
Moreover, it has comparable threshold with the best renormalization
group decoders \cite{duclos2010renormalization}. For code-sizes up
to 200 physical qubits the decoder is practical and we discuss how
to extend our neural network architecture to negate the inefficiencies
that kick in at that stage.

Machine learning techniques, specifically neural networks, have been
gaining popularity over the last year, in particular with the recent
developments in using restricted Boltzmann machines for describing
the ground state of many-body systems \cite{carleo2017solving} or
convolutional networks for identifying phases of matter \cite{carrasquilla2016machine}.
A preprint on the use of restricted Boltzmann machines to decoding
the toric code has been available for a few months as well \cite{torlai2016neural},
however that architecture does not yet outperform known decoders like
MWPM and has been tested only on the Z syndrome on lattices no bigger
than 5-by-5. At the time of submission of this manuscript two other
related preprints were made available: a fast neural network decoder
for small surface codes, that however also does not significantly
outperform MWPM \cite{varsamopoulos2017decoding}, and a recurrent
neural network decoder outperforming MWPM as evaluated on a 17 qubit
surface code \cite{baireuther2017machine}. It is worth noting as
well that over the last few months work has started on deep learning
methods for decoding classical algebraic codes \cite{nachmani2016learning}.

\begin{figure}
\centering{}\includegraphics[width=8cm]{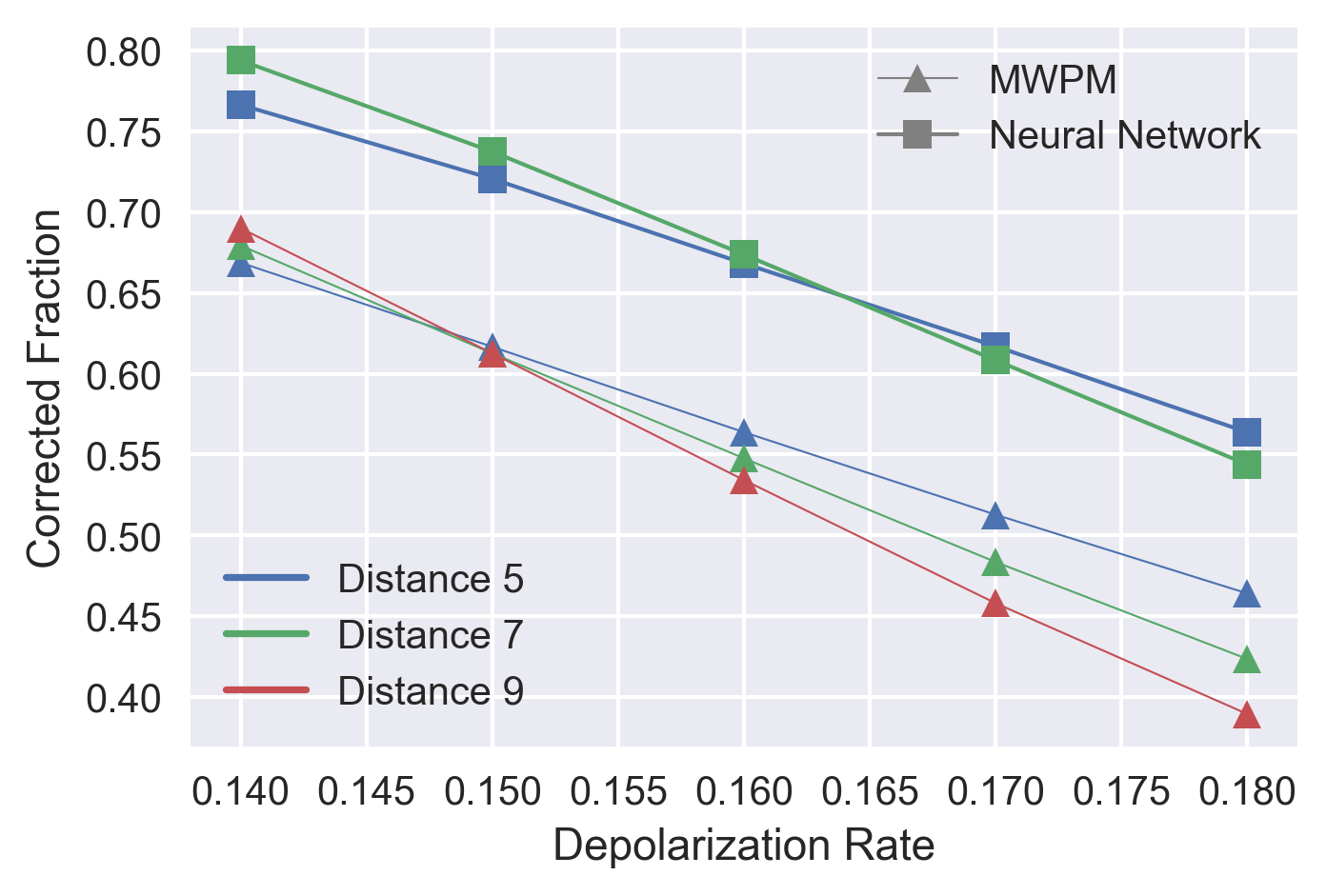}\caption{\textbf{\label{fig:threshold}Decoder performance for toric codes
of distances 5 and 7.} The x axis is the depolarization rate of the
physical qubits (the probability that an X, Y, or Z error has occurred),
while the y axis is the fraction of properly decoded code iterations
(the conjugate of the logical error rate). The neural network decoder
(rectangular markers) significantly outperforms the minimal weight
perfect matching decoder (triangular markers), both in terms of threshold
and logical error rate. For the above plots, neural networks with
18 hidden layers were used.}
\end{figure}

\section*{Results}

For testing purposes we trained our neural decoder (depicted in \figref{ToricCodeDecoder}b)
on the toric code, which already has a number of known decoding algorithms
specifically tuned for its lattice structure. The evaluation was done
under the depolarization error model. Our algorithm significantly
outperforms the standard MWPM decoder. The comparison of the two decoders
in \figref{threshold} shows a threshold single-qubit error which
is nearly 2 percentage points higher for the new algorithm (around
16.4\% for the depolarization error model), and the fraction of correctly
decoded errors is consistently around 10 percentage points higher
than the fraction of errors corrected by MWPM. Furthermore, the neural
decoder threshold compares favorably to renormalization group decoders
\cite{duclos2010renormalization} (threshold of 15.2\%), and decoders
explicitly tuned to correlations between Z and X errors \cite{delfosse2014decoding}
(threshold of 13.3\% for a triangular lattice, as it is tuned for
asymmetric codes). To our knowledge only a renormalization group decoder
\cite{duclos2010fast} enhanced by a sparse code decoder \cite{poulin2008iterative}
reaches a similar threshold (16.4\%). It is worth noting that the
sampling procedure in our decoder makes it impractically slow for
codes of more than 200 physical qubits, while other decoders remain
practical. On the other hand, the neural architecture is versatile
enough to be applied to any stabilizer code, unlike the other decoders
discussed here, which are limited to only topological codes. The best
of both worlds \textemdash{} record threshold and fast decoding \textemdash{}
should be achievable if we couple the renormalization decoder of \cite{duclos2010fast}
with our neural decoder (instead of the currently suggested sparse
code decoder \cite{poulin2008iterative}), however this will be applicable
only to topological codes. We discuss other ways to avoid the inefficiencies
in our decoder without compromising its ability to ``learn'' to
decode any stabilizer code. 

After being properly trained for a given error rate of a particular
error model, the neural network at the heart of our decoder becomes
a compact approximate representation of the probability distribution
of errors that can occur. The decoding algorithm consist of inputing
the measured syndrome in the neural network, interpreting the output
as a probability distribution of the errors conditioned on the given
syndrome, and repeatedly sampling from that distribution. The performance
of the decoder scales monotonically with the size of the network,
up to a point of diminishing returns where using more than about 15
hidden layers (for a distance 5 code) stops providing improvements.

\begin{figure}
\centering{}\includegraphics[width=8cm]{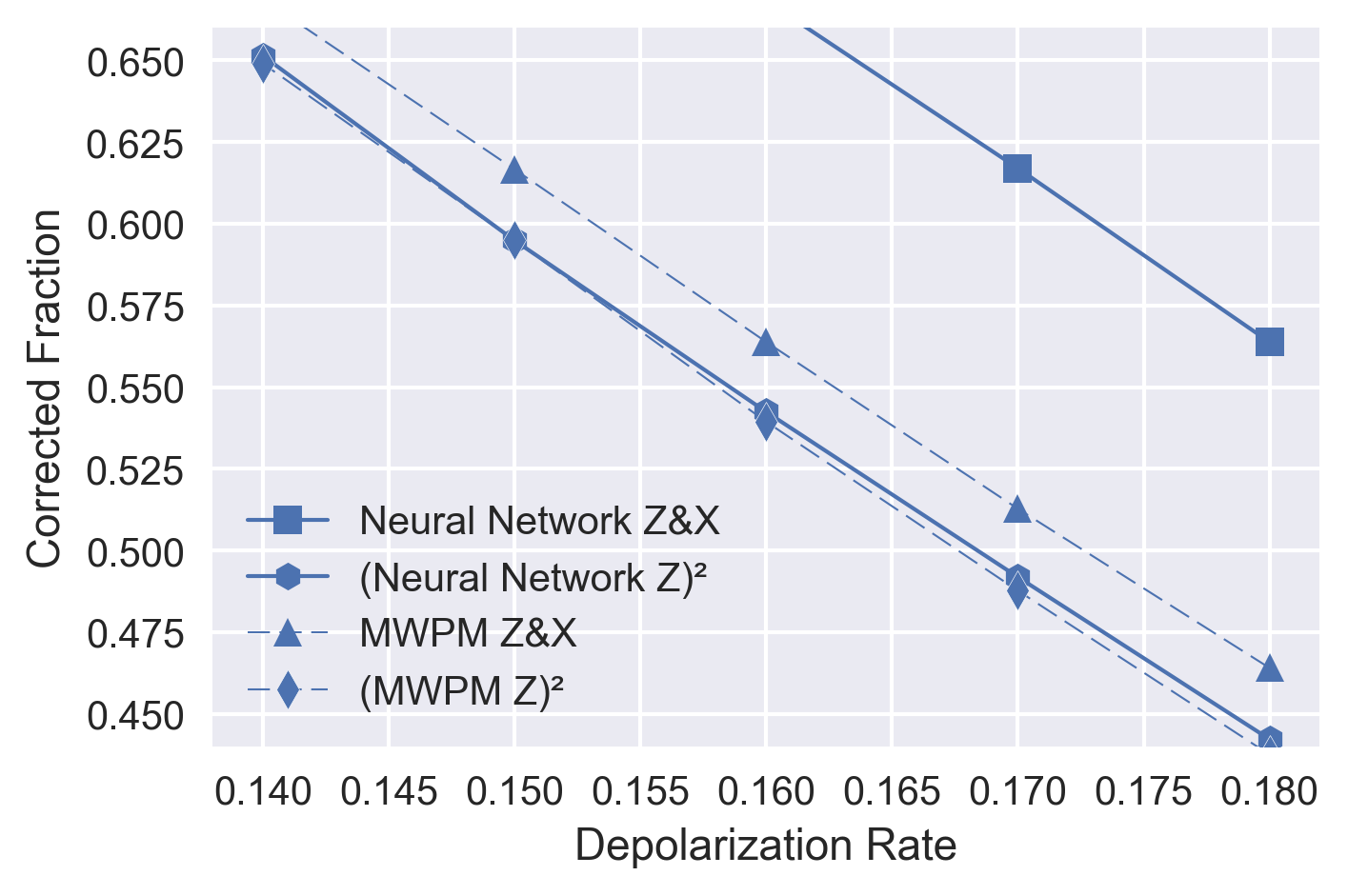}\caption{\textbf{\label{fig:correlations}Correlations learned by the neural
network.} The neural network and MWPM decoder performances for a distance
5 code from \figref{threshold} are repeated in this plot. To visualize
the importance of taking into account correlations between errors,
we also plot the square of the ``corrected fraction'' for a neural
and a MWPM decoder decoding only the Z stabilizer (this neural decoder
was trained only on Z stabilizer data). The neural decoders outperforms
MWPM both when decoding only the Z stabilizer and when decoding Z
and X together. If there were no correlations between errors then
the squared value for decoding Z would be the same as the value for
decoding both Z and X for each decoder. However, the difference is
much more substantial between the two neural decoders, demonstrating
the limitations of MWPM and similar decoders that do not account for
correlations.}
\end{figure}

The significant gain in the threshold value relative to some known
decoders can be traced to two characteristics of the neural network
(discussed in more details in the Methods section). Firstly, the neural
network is trained on (stabilizer, error) pairs generated from the
error model, therefore it is optimized directly for producing ``most
probable error'', not for finding an imperfect proxy like ``error
with lowest energy'' as is the case for MWPM. Secondly (depicted
in \figref{correlations}), it learns the Z and X stabilizers together,
hence it can encode correlations between them in its structure. Namely,
in a typical depolarization error models, one third of the errors
are Y errors (equivalent to both X and Z error happening), therefore
the knowledge of this correlation can be a useful resource for decoding.
Other decoders need significant modifications to even partially employ
those correlations in decoding \cite{delfosse2014decoding}.

\section*{Methods}

Neural networks are particularly efficient tools for function approximation
\cite{cs231n}, where a function $f:\ x\rightarrow f(x)$ is to be
learned from large amount of training data given in the form of pairs
$(x,\ f(x))$. The input $x$ is set as the value of the input layer
of neurons. Each of those neurons is connected through axons with
each neuron of the next layer (the first ``hidden'' layer). Multiple
hidden layers of neurons can be connected together in this fashion
in order to construct a deeper neural network. The last layer of the
network is the output layer - its value represents $f_{learned}(x)$.
The value of a neuron (i.e. its activation value) is calculated as
a weighted sum of the activation values of the neurons connected to
it from the previous layer. That sum is then passed through a non-linear
function (called the activation function). This activation value is
then further passed on to the neurons of the next layer, where the
process is repeated until it reaches the output layer. The weights
in the sums (i.e. the strength of connections between the neurons)
are parameters which are optimized through stochastic gradient descent
in order to minimize the distance between $f_{learned}$ and $f$
calculated on the training data. The choice of activation function,
the size of the hidden layers, and the step size for gradient descent
(also called the hyperparameters) are decided in advance, before training.
Current best practices include performing a random search to find
the best hyperparameters.

In the particular case of decoding a stabilizer quantum error correcting
code we want to map syndromes to corresponding physical errors, hence,
we take the input layer to be the syndrome (obtained from measuring
the stabilizers). For instance, for a toric code of lattice size 9-by-9
we have to measure 81 plaquette operators and 81 star operators for
a total of 162 input neurons (having value 0 if the syndrome is trivial
and 1 if not). Similarly, we set the output layer to be the prediction
for what physical errors occurred (typically represented in the Heisenberg
picture, thanks to the Gottesman\textendash Knill theorem). Using
the same example, we have 162 physical qubits and we need to track
their eigenvalues under both Z and X operators, requiring a total
of 324 output neurons (having value 0 if no error has occurred and
value 1 otherwise).

To completely define the neural network architecture we set the activation
functions of the hidden layers to $\tanh$ and the activation of the
output layer to the sigmoid function $\sigma(x)=\left(1-e^{-x}\right)^{-1}\in[0,1]$.
The size of the hidden layer was set to four times the size of the
input layer. These decisions were reached after an exhaustive search
over possible hyperparameters tested on toric codes of distance 3
to 6, and proved to work well for bigger codes as well. The number
of hidden layers was varied - deeper networks produce better approximations
up to a point of diminishing returns around 15 layers. The step size
for the gradient descent (a.k.a. the learning rate) was annealed -
gradually lowered, in order to permit rapidly reaching the minimum.
The distance measure between training and evaluation data that is
being minimized by the gradient descent is their binary crossentropy
(a measure of difference between two probability distributions discussed
below).

The training was done over one billion (syndrome, error) pairs in
batches of 512, taking about a day of GPU wall time for a 5-by-5 toric
code. The pairs were generating on the fly, by first generating a
sample error from the given error model (this training set can also
be efficiently generated directly on the experimental hardware), and
then obtaining the corresponding syndrome by a dot product with the
parity check matrix. The error model used for each physical qubit
was qubit depolarization, parametrized by qubit fidelity $p$ - the
probability of no error happening on a given qubit, or equivalently
depolarization rate $1-p$. Under this model, Z, X, and Y (consecutive
Z and X) errors had equal probabilities of $\nicefrac{1}{3}(1-p)$.
For each value of p we trained a new network, however the results
showed some robustness to testing a neural network at an error rate
different from the one at which it was trained.

The performance of the network was improved if we normalize the input
values to have an average of 0 and a standard deviation of 1. For
a depolarization error rate $1-p$, the rate at which a Z eigenvalue
flips is $P_{e}=\nicefrac{2}{3}(1-p)$ and independently the rate
for X flips is the same. In the example of the toric code the rate
of non-trivial stabilizer measurements will be the same for Z and
for X, namely $P_{s}=4q^{3}(1-q)+4q(1-q)^{3}$ and the variance will
be $V_{s}=P_{s}-P_{s}^{2}$.

\begin{figure}
\centering{}\includegraphics[width=7cm]{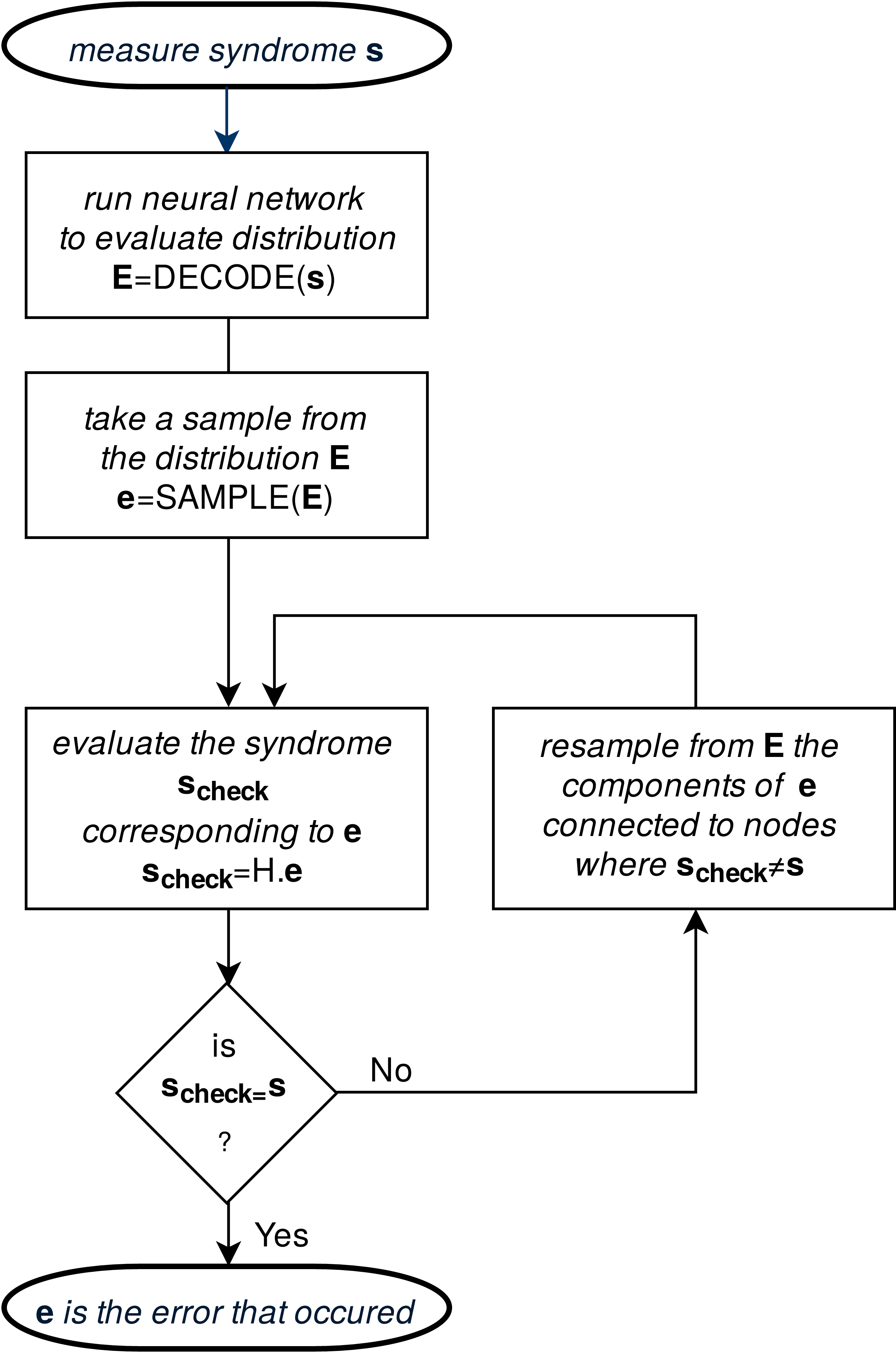}\caption{\textbf{\label{fig:Sampling-the-neural-network}Sampling the neural
network.} (Arrays in the diagram are in bold font, $H$ is the parity
check matrix of the code) After the neural network is trained its
output can efficiently be evaluated for any given syndrome $\boldsymbol{s}$.
The output array $\boldsymbol{E}$ is interpreted as a list of probabilities
for each qubit for an error to have happened. An array $\boldsymbol{e}$
(whether an error occurred at each qubit) is sampled from $\boldsymbol{E}$.
In the loop we check whether the guess $\boldsymbol{e}$ actually
produces the same syndrome as the initially given one. If not, we
resample only the qubits taking part in the stabilizer measurement
corresponding to the incorrect elements of the syndrome. If the loop
runs for more than a set number of iterations we declare failure to
decode (a detected, however not corrected, error). As for any other
decoding algorithm the final result may be wrong if the total error
that occurred is of particularly low probability (i.e. of high weight).
In the case of a general stabilizer code $H$ stands for the list
of stabilizer operators. }
\end{figure}

At this point we have not discussed yet how to use the fully trained
neural network in decoding. A trained network can efficiently evaluate
the approximation of the decoding function (from here on referred
to as $\text{DECODE}:\text{syndrome}\rightarrow\text{error}$), so
all Alice needs to do in order to perform error correction on her
quantum memory is to measure the syndrome and run the neural network
forward to evaluate $\text{DECODE(syndrome)}$. However, the neural
network is a continuous function and an imperfect approximation, therefore
the values in $\text{DECODE(syndrome)}$ will not be discrete zeros
and ones, rather they will be real numbers between zero and one. A
common way to use and interpret those values is to view them as a
probability distribution over possible errors, i.e. the i-th value
in the array $\text{DECODE(syndrome)}$ is a real number between zero
and one equal to the probability of the i-th qubit experiencing a
flip (half of the array corresponds to Z errors and half of the array
corresponds to X errors). This interpretation is reinforced by our
use of binary crossentropy as an optimization target during training.
In order to deduce what error has occurred we sample this probability
distribution. We verify the correctness of the sample by computing
the syndrome that the predicted error would cause - if it differs
from the given syndrome we resample. This sampling procedure is present
in \cite{torlai2016neural} as well, however we further employ a simple
``hard decision belief propagation / message passing'' sampling,
which can speed up the sampling process by an order of magnitude:
we resample only the qubits taking part in the stabilizer measurement
corresponding to the incorrect elements of the syndrome (see \figref{Sampling-the-neural-network}).

\paragraph*{Data Availability}

The code for building, training, and evaluating the neural network
decoder is publicly available on the authors' web page, and shell
scripts with the parameters for the presented figures are available
upon request. Pretrained neural networks can be provided as well.

\section*{Discussion}

On first sight our decoder implementation can look like a look-up
table implementation, however we would like to stress the immense
compression of data that the neural network achieves. Firstly, one
can consider the size of the neural network itself. For a code on
$N$ physical qubits the number of parameters needed to describe a
neural decoder of $L$ layers will be $\mathcal{O}\left(N^{2}L\right)$
or on the order of thousands for the codes we tested. Moreover, the
size of the training dataset for the codes we tested did not exceed
10 billion, and it can be made orders of magnitude smaller if we reuse
samples in the stochastic gradient descent (a common approach taken
in training). On the other hand, the size of a complete lookup table
would be on the order of $\mathcal{O}\left(4^{N}\right)$. Even if
we take only the most probable errors (and discard the errors that
have less than 5\% chance of occurring), at depolarization rate of
$0.1$ we need a lookup table bigger than $10^{12}$ for a distance
5 toric code (50 qubits), bigger than $10^{23}$ for distance 7 toric
code (98 qubits), and bigger than $10^{37}$ for distance 9 toric
code (162 qubits).

Thanks to this compression, to the direct optimization for most probable
error, and to the ease of including knowledge of error correlations
in the decoding procedure, the algorithm presented here is one of
the best choices for decoding stabilizer codes of less than 200 qubits.
While we used the toric code for our testing, there is nothing in
our design that has knowledge of the specific structure of that code
- the neural decoder can be applied to the decoding of any stabilizer
code.

\begin{figure}
\centering{}\includegraphics[width=8cm]{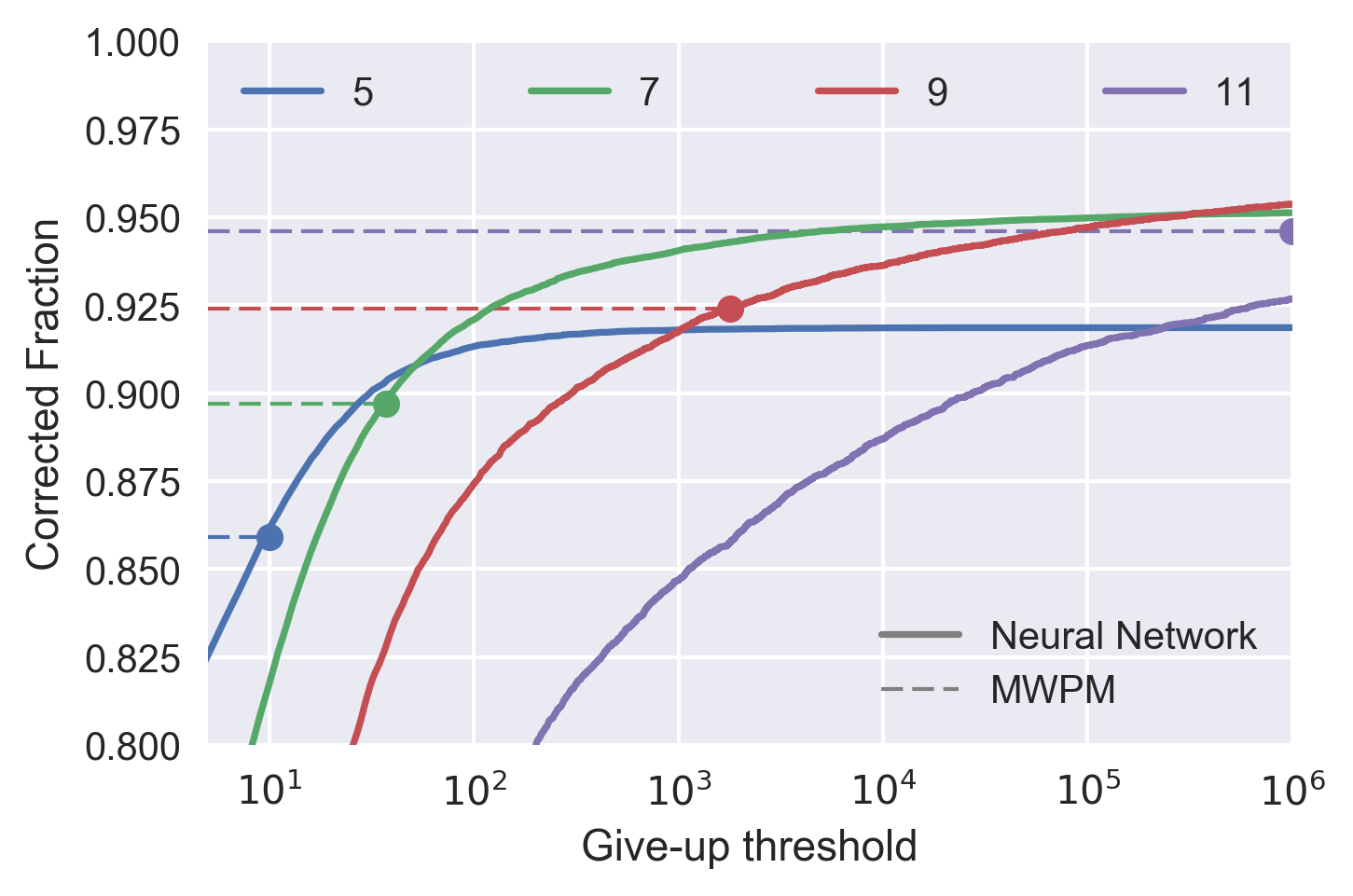}\caption{\textbf{\label{fig:Sampling-giveup}Sampling overhead versus decoder
performance.} Sampling possible errors from the output of the neural
network is an iterative process not guaranteed to reach an acceptable
solution, therefore we need to set an upper bound on how many iterations
are permitted before giving up (which would result in a detected but
not corrected error). The plot gives the performance of our decoder
trained on toric codes of different distances with respect to the
maximal permitted number of iterations. The dashed lines give the
MWPM decoder performances for the same codes as a reference. Codes
up to distance 9 (containing 162 physical qubits) are practical, but
extending using our decoder for codes with more than 242 physical
qubits would be prohibitive due to the sampling overhead. The evaluations
were done for 10\% depolarization rate on the physical qubits.}
\end{figure}

Due to the probabilistic nature of the sampling, the decoder becomes
impractically inefficient for codes bigger than roughly 200 qubits
as one can see in \figref{Sampling-giveup}. This can be attributed
to two characteristics of our algorithm: we use a simple hard-decision
message passing algorithm in our sampling instead of a more advanced
belief propagation algorithm seeded by output of the neural network;
additionally, our neural network learns only the marginal probabilities
for errors on each qubit, without providing the correlations between
those errors. A more advanced neural network could address this problem
by providing correlation information in its output layer. Our focus
forward goes beyond that: we can consider recurrent generative networks
\cite{rumelhart1986learning} that have the belief propagation as
part of their recurrent structure.

While this decoder is general and it can be applied to any stabilizer
code, one can also design neural network architectures that specifically
exploit the lattice structure and translational symmetry of the toric
code. For instance, convolutional neural networks are well adapted
for processing 2D data. Moreover thanks to the translational symmetry
one can envision a decoder that is trained on a fixed patch of the
code and it can be used for toric codes of any size. As already mentioned,
our decoder can readily replace the sparse code decoder \cite{poulin2008iterative}
used as part of the renormalization group decoder of \cite{duclos2010fast},
hence providing great decoding speed and high threshold values.

\section*{Acknowledgements}

We acknowledge the stimulating discussions with Kasper Duivenvoorden,
Steven Flammia, Steven Girvin, Alexandar Makelov, and Mehmet Tuna
Uysal. We thank the Yale HPC staff for the provided computing resources.
We acknowledge support from the ARL-CDQI (W911NF-15-2-0067), ARO (W911NF-14-1-0011,
W911NF-14-1-0563), ARO MURI (W911NF-16-1-0349 ), AFOSR MURI (FA9550-14-1-0052,
FA9550-15-1-0015), NSF (EFMA-1640959), Alfred P. Sloan Foundation
(BR2013-049), and Packard Foundation (2013-39273).

\subsubsection*{Additional Information}

SK contributed the code for the project. Design, analysis, and manuscript
preparation was contributed jointly by SK and LJ.

Competing Financial Interests statement: There is no competing interest.

\bibliographystyle{unsrt}
\bibliography{neural_decoder}

\begin{thebibliography}{10}

\bibitem{Shannon48}
Claude~E Shannon.
\newblock The mathematical theory of communication.
\newblock {\em The Bell System Technical Journal}, 27:379--423, 623--656, 1948.

\bibitem{Lidar13}
Daniel~A. Lidar and eds. Todd A.~Brun.
\newblock {\em Quantum Error Correction}.
\newblock Cambridge University Press, 2013.

\bibitem{Terhal15}
Barbara~M. Terhal.
\newblock Quantum error correction for quantum memories.
\newblock {\em Rev. Mod. Phys.}, 87(2):307--346, 2015.

\bibitem{von1956probabilistic}
John Von~Neumann.
\newblock Probabilistic logics and the synthesis of reliable organisms from
  unreliable components.
\newblock {\em Automata studies}, 34:43--98, 1956.

\bibitem{NC00}
M.~A. Nielsen and Isaak Chuang.
\newblock {\em Quantum computation and quantum information}.
\newblock Cambridge University Press, Cambridge, U.K; New York, 2000.

\bibitem{gottesman1997stabilizer}
Daniel Gottesman.
\newblock Stabilizer codes and quantum error correction.
\newblock {\em arXiv:quant-ph/9705052}, 1997.

\bibitem{calderbank1996good}
A~Robert Calderbank and Peter~W Shor.
\newblock Good quantum error-correcting codes exist.
\newblock {\em Physical Review A}, 54(2):1098, 1996.

\bibitem{Shor96}
Peter~W. Shor.
\newblock Fault-tolerant quantum computation.
\newblock In {\em Proc. 37nd Annual Symposium on Foundations of Computer
  Science}, pages 56--65. IEEE Computer Society Press, 1996.

\bibitem{steane1997active}
Andrew~M Steane.
\newblock Active stabilization, quantum computation, and quantum state
  synthesis.
\newblock {\em Physical Review Letters}, 78(11):2252, 1997.

\bibitem{gallager1962low}
Robert Gallager.
\newblock Low-density parity-check codes.
\newblock {\em IRE Transactions on information theory}, 8(1):21--28, 1962.

\bibitem{MacKay96}
David~JC MacKay and Radford~M Neal.
\newblock Near shannon limit performance of low density parity check codes.
\newblock {\em Electron. Lett.}, 32(18):1645, 1996.

\bibitem{poulin2008iterative}
On the iterative decoding of sparse quantum codes.
\newblock {\em Quantum Information and Computation}, 8:987--1000, 2008.

\bibitem{Kitaev03}
A.~Yu Kitaev.
\newblock Fault-tolerant quantum computation by anyons.
\newblock {\em Annals of Physics}, 303(1):2--30, 2003.

\bibitem{dennis2002topological}
Eric Dennis, Alexei Kitaev, Andrew Landahl, and John Preskill.
\newblock Topological quantum memory.
\newblock {\em Journal of Mathematical Physics}, 43(9):4452--4505, 2002.

\bibitem{Edmonds65}
Jack Edmonds.
\newblock Paths, trees, and flowers.
\newblock {\em Canadian Journal of mathematics}, 17(3):449--467, 1965.

\bibitem{duclos2010renormalization}
Guillaume Duclos-Cianci and David Poulin.
\newblock A renormalization group decoding algorithm for topological quantum
  codes.
\newblock In {\em Information Theory Workshop (ITW), 2010 IEEE}, pages 1--5.
  IEEE, 2010.

\bibitem{carleo2017solving}
Giuseppe Carleo and Matthias Troyer.
\newblock Solving the quantum many-body problem with artificial neural
  networks.
\newblock {\em Science}, 355(6325):602--606, 2017.

\bibitem{carrasquilla2016machine}
Juan Carrasquilla and Roger~G Melko.
\newblock Machine learning phases of matter.
\newblock {\em Nature Physics}, 13:431--434, 2017.

\bibitem{torlai2016neural}
Giacomo Torlai and Roger~G Melko.
\newblock Neural decoder for topological codes.
\newblock {\em Physical Review Letters}, 119(3):030501, 2017.

\bibitem{varsamopoulos2017decoding}
Savvas Varsamopoulos, Ben Criger, and Koen Bertels.
\newblock Decoding small surface codes with feedforward neural networks.
\newblock {\em arXiv:1705.00857}, 2017.

\bibitem{baireuther2017machine}
P~Baireuther, TE~O'Brien, B~Tarasinski, and CWJ Beenakker.
\newblock Machine-learning-assisted correction of correlated qubit errors in a
  topological code.
\newblock {\em arXiv:1705.07855}, 2017.

\bibitem{nachmani2016learning}
Eliya Nachmani, Yair Beery, and David Burshtein.
\newblock Learning to decode linear codes using deep learning.
\newblock {\em arXiv:1607.04793}, 2016.

\bibitem{delfosse2014decoding}
Nicolas Delfosse and Jean-Pierre Tillich.
\newblock A decoding algorithm for css codes using the x/z correlations.
\newblock In {\em Information Theory (ISIT), 2014 IEEE International Symposium
  on}, pages 1071--1075. IEEE, 2014.

\bibitem{duclos2010fast}
Guillaume Duclos-Cianci and David Poulin.
\newblock Fast decoders for topological quantum codes.
\newblock {\em Physical review letters}, 104(5):050504, 2010.

\bibitem{cs231n}
Andrej Karpathy.
\newblock Stanford university {CS231n}: Convolutional neural networks for
  visual recognition.
\newblock 2015.

\bibitem{rumelhart1986learning}
DE~Rumelhart, G~Hinton, and R~Williams.
\newblock Learning sequential structure in simple recurrent networks.
\newblock {\em Parallel distributed processing: Experiments in the
  microstructure of cognition}, 1, 1986.

\end{thebibliography}

\pagebreak{}

\appendix

\section*{Supplementary Materials}

\subsubsection*{Performance of sampling through ``hard-decision message passing''}

In the main text we mention that sampling from the neural network
can be done in two different manners: either through naive resampling
in which the entire ``candidate error vector'' sample is scraped
if it does not reproduce the measured syndrome or through a more advanced
``hard-decision message passing'' which provides for a significant
speedup. The message passing works by checking which qubits belong
to the violated syndrome components (i.e. which variable nodes are
connected to the active check nodes on the Tanner graph) - in the
resampling step, only those qubits are resampled, hence preserving
the already properly decoded components of the error vector. The following
figure, similar to the figures in the main text, demonstrates the
speedup for a resampling performed on a deep network decoding the
5x5 code. In order to also compare the hard-decision message sampling
on its own, without the use of a neural network (i.e. the way message
passing is currently used in classical LDPC decoders), we also show
a curve where the message passing sampler is run on an untrained network
(as expected, it performs poorly - it is known that message passing
on its own does not work well for quantum LDPC codes due to the presence
of 4-cycles in the Tanner graph).
\begin{center}
\includegraphics[width=8cm]{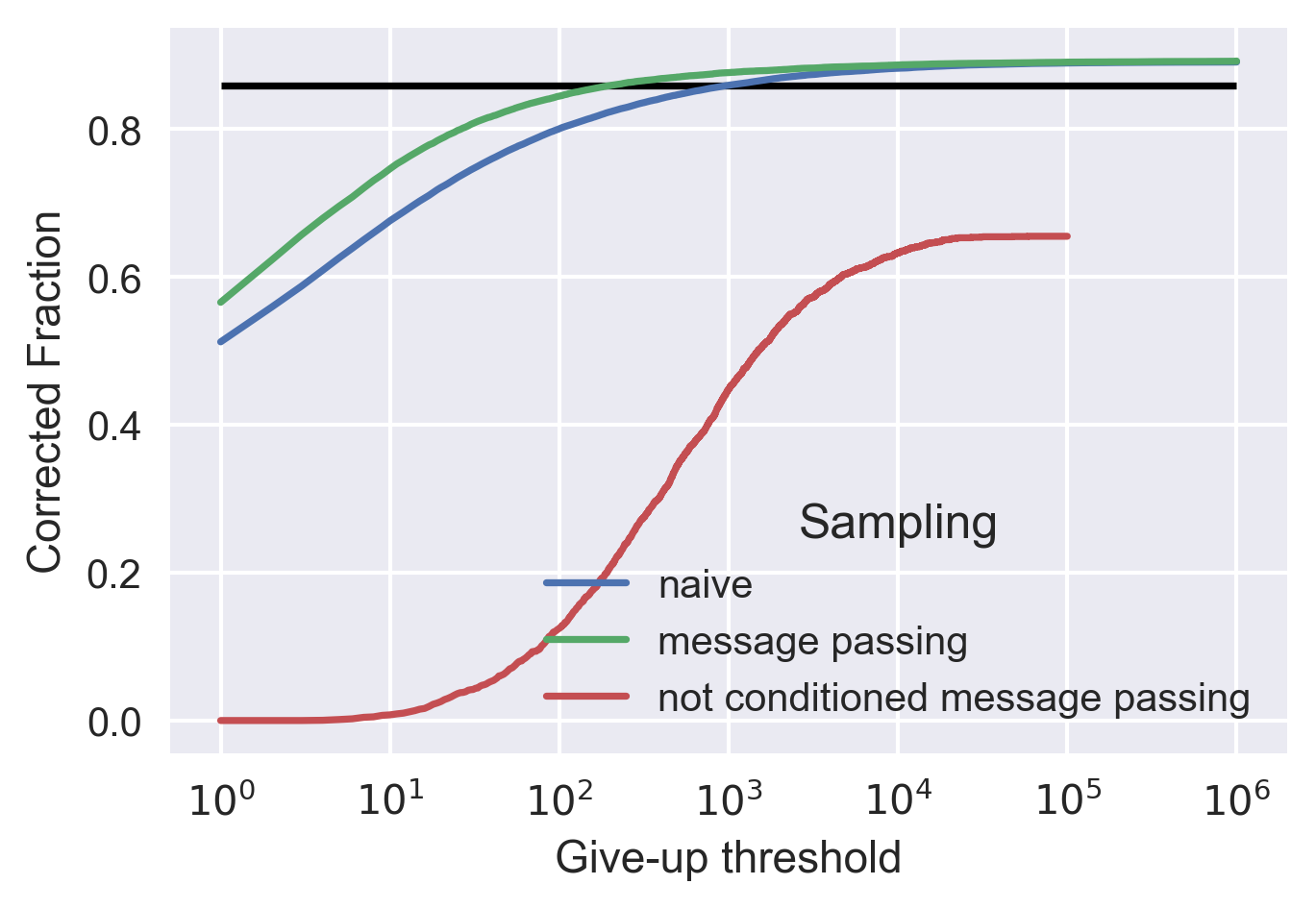}
\par\end{center}

\subsubsection*{Performance vs depth of the neural network}

The following figure, similar in style to figures in the main text,
shows the growth in performance as we increase the depth of the neural
network (the example is for a 5x5 toric code). As discussed, a point
of diminishing returns is reached, where the expense of adding more
layers outweighs the minor gains in performance.
\begin{center}
\includegraphics[width=8cm]{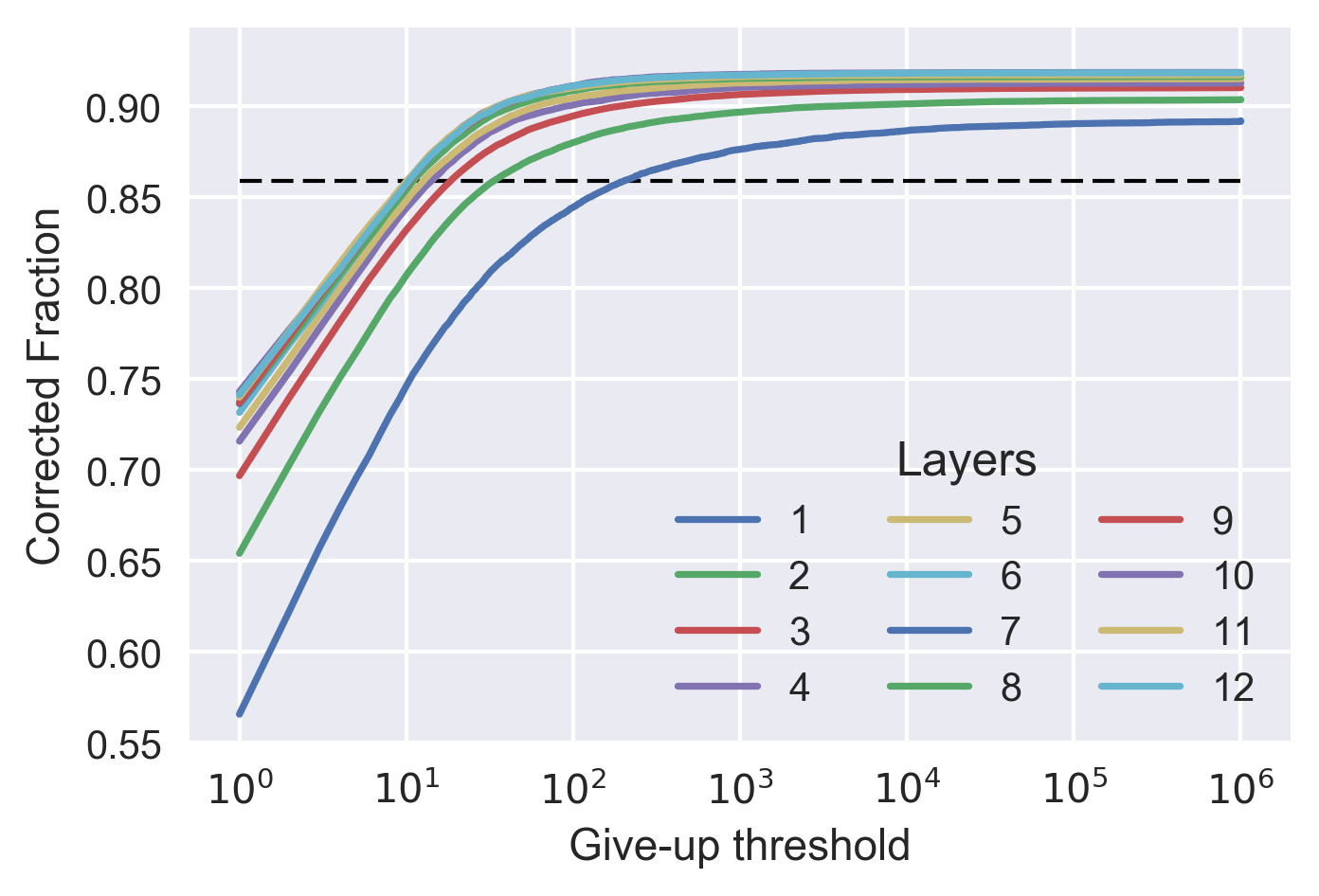}
\par\end{center}

\subsubsection*{Software package}

The software described in the main text is available from the authors.
It is based on the Keras and Tensorflow NN libraries and can run on
GPU accelerators. The software provides independent command line utilities
that can be used to design and evaluate deep neural network decoders
with arbitrary hyperparameters. Upon request we would be happy to
provide pretrained networks for any reasonably sized code of interest.
\end{document}